TiO$_2$ as an electrostatic template for epitaxial growth of EuO on MgO(001) by reactive molecular beam epitaxy

A. G. Swartz[†], Jared J. I. Wong[†], I. V. Pinchuk, and R. K. Kawakami*

Department of Physics and Astronomy, University of California, Riverside, CA 92521

Abstract

We investigate the initial growth modes and the role of interfacial electrostatic interactions of EuO epitaxy on MgO(001) by reactive molecular beam epitaxy. A TiO$_2$ interfacial layer is employed to produce high quality epitaxial growth of EuO on MgO(001) with a 45° in plane rotation. For comparison, direct deposition of EuO on MgO, without the TiO$_2$ layer shows a much slower time evolution in producing a single crystal film. Conceptual arguments of electrostatic repulsion of like-ions are introduced to explain the increased EuO quality at the interface with the TiO$_2$ layer. It is shown that ultrathin EuO films in the monolayer regime can be produced on the TiO$_2$ surface by substrate-supplied oxidation and that such films have bulk-like magnetic properties.

† These authors contributed equally to this work.

* email: roland.kawakami@ucr.edu

1. Introduction

The spin filter effect,[1] possible use as a magnetic gate dielectric,[2,3] and a large magneto-optic response,[4] makes stoichiometric EuO, a ferromagnetic insulator, promising for spin-based applications.[5] Also of great interest are doped and nonstoichiometric EuO due to their demonstration of a metal-insulator transition,[6] colossal magnetoresistance,[7] half metallic behavior,[8] and the anomalous Hall effect.[9] The recent resurgence of interest in EuO is largely due to the advances in synthesis of high quality EuO films by reactive molecular beam epitaxy (MBE).[3,10-12] More specifically, the stoichiometric growths have been reliably achieved only within an "adsorption-controlled" growth regime.[10,11] Two separate conditions determine this regime. First, the substrate is maintained at an elevated temperature, which allows for Eu re-evaporation (distillation) from the substrate. Second, a carefully maintained oxygen partial pressure determines the growth rate and chemical composition ($Eu_xO_y$).

MgO is an important oxide for spintronics due to its $\Delta_1$ band spin filtering in magnetic tunnel junctions[13-15] and its effective use as a tunnel barrier for spin injection into semiconductors and graphene.[16-19] Also, MgO has long served as a popular commercially available substrate for the deposition of a wide variety of materials such as transition metals, perovskites, and spinels.[13,20,21] Several authors have reported successful deposition of EuO on MgO[12,22] and cube-on-cube growth with a magnetization of 7 Bohr magnetons per Eu atom despite the large lattice mismatch of ~22% (($a_{EuO}-a_{MgO}$) / $a_{MgO}$ = (0.514 nm − 0.421 nm) / 0.421 nm = 22.1%).[3] However, while single crystal deposition on MgO(001) is possible, the initial stages of the growth have yet to be fully investigated and require further exploration.[12,23]

In this article we present the results of high quality EuO epitaxy on MgO by the introduction of a $TiO_2$ interfacial layer. Conceptual electrostatic arguments are introduced to explain why $TiO_2$ alleviates many of the problems associated with rock salt heteroepitaxy. Time evolution of the growths are compared and the $TiO_2$ surface is shown to produce single crystal EuO in the monolayer regime by inducing a 45° in plane rotation, which decreases the lattice mismatch, and by serving as an electrostatic template for which like-ion repulsion is alleviated. On the other hand, direct epitaxy of EuO on MgO is shown to be of reasonable quality only after 2nm. Interestingly, ultrathin EuO can be produced without the introduction of oxygen partial pressure through substrate-supplied oxidation to yield films in the monolayer regime. Such ultrathin films are ferromagnetic with bulk Curie temperatures.

2. Experimental Details

In this study, 10mm x 10mm x 0.5mm double-side polished MgO(001) substrates are first rinsed in DI water, then loaded into a MBE system with a base pressure ~1x10$^{-10}$ torr. The crystal surface quality of the sample is monitored throughout the annealing and subsequent layer growths with *in situ* reflection high energy electron diffraction (RHEED). The substrate is annealed for 60 minutes at 600°C as measured by a thermocouple located near the sample. The substrate is then cooled to 350°C for the deposition of a 10 nm MgO buffer layer grown by e-beam evaporation at a typical rate of ~1 Å/min.[24] The MgO buffer layer smoothes the substrate's surface, indicated in the RHEED pattern as sharpened streaks and Kikuchi lines (Fig 2 (A) and 2 (B)). To create the TiO$_2$ layer, Ti is first deposited from an e-beam source onto the MgO buffer layer at room temperature (RT). The Ti thickness is chosen according to the number of desired surface Ti atoms corresponding to 1, 1.5, or 2 monolayers of lattice matched 2x2 reconstructed TiO (chemical composition TiO$_2$) as described more fully in the following section. The Ti layer is exposed to molecular oxygen (5x10$^{-8}$) at 500°C for 30 minutes. For subsequent growths on either the TiO$_2$ or directly on the MgO buffer layer, EuO films are produced by reactive MBE where a high purity metal source is sublimed and allowed to react with a molecular oxygen partial pressure. Typical stoichiometric growth in the adsorption-controlled (distillation and oxygen-limited) regime proceeds as follows. 99.99% pure Eu metal is evaporated from a thermal effusion cell and the flux (~8 Å/min) is incident upon the heated substrate which is maintained at 500°C. Next, molecular oxygen is leaked into the chamber with a partial pressure of 1x10$^{-8}$ torr enabling the growth of stoichiometric EuO.[3,11,12] Such films on bare MgO have been shown to be approximately 5nm thick for a 30minute growth time by AFM profiling giving a growth rate of 0.17 nm/min.[3]

3. Growth and Electrostatics at the EuO/MgO(001) Interface

Heteroepitaxy between insulating oxides, such as of large cation oxides on MgO, is greatly determined by interface electrostatics.[20] Purely structural considerations are insufficient to fully understand the EuO/MgO interface. The cube-on-cube (EuO(001) [100] // MgO(001) [100]) growth on MgO[3,12] suggests that some structural arrangement (i.e. either 1:1, 3:4, 4:5, etc…) is favored. A 3:4 spacing has a reduced lattice mismatch of 8.4% and a 4:5 spacing has a mismatch of 2.3%. For a clearer picture, a 4:5 (EuO:MgO) stacking, displayed using VESTA software,[25] is shown in Figure 1 (A). An examination of the 4:5 stacking shows that while the center $Eu^{2+}$ ion has a favorable position above an $O^{2-}$ ion, at the left edge, the first $Eu^{2+}$ ion is sitting above an $Mg^{2+}$ ion and the first $O^{2-}$ ion is above another $O^{2-}$ ion. This is repeated at the right edge of the 4:5 configuration. From an electrostatic point of view, strong Coulomb repulsion between like ions suggests that such a stacking is not ideal despite the improved lattice match and would certainly lead to surface roughening at the interface. Another concern for the cube-on-cube growth mode for direct heteroepitaxy is the ion-size difference effect,[20] which is related to the difference in size between the Mg-O bond and the Eu-O bond. The $Mg^{2+}$ ionic diameter, 0.130 nm, combined with two Oxygen ($O^{2-}$) ionic radii of 0.140 nm, forms a nearly close-packed system with the ions spanning 97% of the lattice constant.[26] Figure 1 (B) illustrates that replacing the $Mg^{2+}$ ion with a $Eu^{2+}$ ion changes the cation ionic diameter to 0.234 nm and increases the O-O nearest neighbor bond by 22% from 0.298 nm to 0.363 nm. Effectively, the deposition of an atomically flat EuO layer on a pristine MgO(001) surface is equivalent to 100% substitutional doping the Mg atoms in the top layer of an MgO surface with Eu atoms. In such a case, the ion-size difference would force the Eu or O atoms to find equilibrium positions in a roughened structure.

Alternatively, another possible structural alignment would be a 45° in-plane rotation of the EuO lattice relative to the underlying MgO orientation. Figure 1 (C) shows a 45° rotated EuO layer on an MgO underlayer with the placement of a Eu ion on an oxygen bonding site. Such a configuration would remove the ion-size effect ($a_{EuO} / 2 = 2.57 < a_{MgO} / \sqrt{2} = 2.91$), reduce the lattice mismatch to 12% and could potentially improve the growth mode. However, anion-anion or cation-cation electrostatics makes the structure energetically unfavorable because the oxygen ions in the EuO overlayer sit atop oxygen ions in the underlying MgO surface. With these considerations in mind, any attempt to engineer the interface to minimize the electrostatic repulsion of like ions, while simultaneously maintaining an atomically smooth surface, could greatly improve the epitaxy.

To alleviate the interfacial electrostatic repulsion and stabilize EuO epitaxy on MgO(001), we propose a special $TiO_2$ template at the interface. Such an approach has been employed to produce high quality epitaxy of $BaTiO_3$ films on MgO(001).[20] Figure 1 (D) shows the stacking for subsequent depositions of a

TiO$_2$ layer followed by EuO on MgO(001). Starting from left to right in Figure 1 (D) is the MgO buffer layer, followed by a monolayer of TiO$_2$, and lastly, a single unit cell of EuO is shown rotated 45° relative to the MgO in-plane orientation. The displayed MgO lattice spacing is that of bulk MgO and the TiO$_2$ layer is shown lattice matched to the MgO. The EuO is shown with bulk EuO lattice constant. For the single monolayer of TiO$_2$, O atoms are positioned above Mg atoms. The Ti atoms and vacancy positions are located above the O atoms of the MgO layer. This configuration allows for the subsequent EuO layer to be positioned such that the Eu atoms are located above the vacant positions in the TiO$_2$ layer, while the O atoms are located above the Ti atoms. Within this EuO/TiO$_2$/MgO(001) interface, all nearest neighbor ions have opposite charge to produce attractive Coulomb forces for an energetically stable interface. Specifically, there are no O-O or Eu-cation nearest neighbor bonds. Thus, the TiO$_2$ interfacial layer eliminates the ion-size effect and electrostatic problems described for the growth of EuO directly onto MgO(001).

4. Experimental Results and Discussion

Figure 2 shows the RHEED patterns for TiO$_2$ monolayers on MgO(001). Figure 2 (A) and (B) show the 10 nm MgO/MgO(001) buffer layer pattern along the [100] and [110] directions, respectively. Figure 2 (C), (E), and (G) are the RHEED images for oxidized Ti layers of 1, 1.5, and 2 ML along the [100] direction, while Figure 2 (D), (F), and (H) are the corresponding TiO$_2$ monolayers along the [110] direction of MgO. The main features of the oxidized Ti patterns remain that of MgO with a slight broadening of the outer diffraction rods. In the RHEED image of 2 ML TiO$_2$/MgO(001) [100] (Figure 2 (G)) the underlying MgO structure is readily visible with the important addition of inner streaks between the main MgO(001) [100] rods. As discussed previously, the TiO$_2$ layer is comprised of both Ti sites and vacant sites above the underlying oxygen atoms. Thus, the unit cell periodicity is increased to twice the size creating diffraction rods of half-spacing in the [100] direction. Equivalently, this TiO$_2$ layer can be perceived as a TiO rock salt surface of identical unit cell with the MgO lattice, but missing the face-centered Ti atoms.[27] This would then be a 2x2 reconstructed TiO surface producing diffraction streaks inside the MgO [100] rods. However, in no instances were inner streaks seen for the case of 1 ML oxidized Ti. Interestingly, as seen in Figure 2 (D) and (F), inner rods appeared for 1.5 and 2 ML of oxidized Ti along the [110] direction. This suggests decreased periodicity of the 2x2 reconstructed TiO$_2$, possibly from an ordered stacking effect or superstructure causing increased periodicity in the k-space lattice along the [110] direction. The subsequent RHEED patterns of a 5 nm EuO film grown on 2 ML TiO$_2$/MgO(001) in the adsorption-controlled regime are shown in Figures 2 (I) and (J). These final films have an in-plane orientation of EuO(001) [110] // MgO(001) [100] and are thus 45° rotated. Importantly, as Fig 1 (B) and Fig 1 (D) illustrate, the ion-size effect is eliminated since the rotated EuO lattice has a smaller unit cell than the underlying TiO$_2$ template. While, generally, EuO growths on 1 ML TiO$_2$ surfaces resulted in polycrystalline films, deposition on 1.5 ML TiO$_2$ surfaces produced high quality EuO single crystal films of identical growth behavior and evolution to depositions on 2 ML TiO$_2$ (see Fig 3(A) and 3(B)). This is interesting since the 1.5 ML TiO$_2$ RHEED only shows part of the features of seen in the 2 ML RHEED, suggesting that the 1.5 ML TiO$_2$ still has the critical structure of the 2x2 reconstructed TiO layer. Because of this result, and in combination with the desire to keep the TiO$_2$ interface as thin as possible, the 1.5 ML TiO$_2$ layer will be used throughout the remainder of this study.

To further examine the growth of EuO on the TiO$_2$ layer, the time evolution of a line cut across the RHEED pattern is monitored along the MgO(001) [110] in-plane crystal direction over the first 10 minutes of EuO growth. A line cut is obtained by plotting the intensity of the image against the CCD camera's horizontal pixel position and therefore crosses several diffraction rods. The initial line cut of 1.5 ML TiO$_2$ (in MgO(001) [110] direction) is shown at the top of Figure 3 (C). After 20 seconds (dashed line (C1)), the Eu flux is introduced and immediately the RHEED begins to change. After the RHEED pattern is stabilized, oxygen is introduced into the chamber (dashed line (C2)), and the RHEED pattern changes to that of EuO(001) [100]. During this period, the RHEED quickly shifts (1 minute) to that of bulk EuO indicating epitaxy within 1 ML with the introduction of oxygen. The final line cut (1 nm EuO) is shown below the time lapse. Analysis of the final EuO/TiO$_2$ line cut compared to the MgO lattice constant gives a EuO lattice parameter of 0.513 ± 0.006 nm. For comparison, the time evolution for direct deposition of EuO on the MgO buffer layer is shown in Figure 2 (D). As indicated by dashed line (D1), elemental Eu flux is directed onto the MgO(001) substrate held at 500 °C. During this period, the RHEED pattern remains that of MgO, indicating that Eu is re-evaporating and not bonding to the surface. Once oxygen is leaked into the system (dashed line (D2)), the time evolution of the RHEED pattern consists of a fading out of the MgO(001) [110] pattern followed by a gradual recovery to a EuO(001) [110] pattern over several minutes (~2 nm). The diffraction rods increase in intensity over the subsequent 20 minutes of the growth.

Several key differences are immediately apparent between the two growths. First, comparative analysis of the diffraction pattern peak positions in the final line cuts between EuO/TiO$_2$/MgO and EuO/MgO demonstrates that EuO epitaxy on the TiO$_2$ is rotated 45-degree in-plane with respect to MgO, while the direct growth on MgO is cube-on-cube. Second, the evolution from the initial line cut to single crystal EuO takes places at a faster rate for deposition on TiO$_2$/MgO and indicates fast strain relaxation for 45° rotated EuO in agreement with observations of EuO growth on Ni.[23] Third, during the distillation period, before the introduction of an oxygen partial pressure, the re-evaporation for each surface is distinctly different. While in both cases the opening of the Eu shutter decreases the RHEED intensity, on bare MgO buffer layer, the incident Eu flux re-evaporates leaving the MgO(001) [110] RHEED pattern unaltered. However, on the TiO$_2$, the incident flux only re-evaporates after an initial time period for which the inner diffraction streaks associated with the TiO$_2$ are lost but the overall MgO diffraction positions in the RHEED pattern are maintained. Lack of bonding and full re-evaporation at 500 °C on bare MgO suggests, in agreement with the discussion in section 3, that there is some additional interfacial energy at the

EuO/MgO interface that inhibits Eu bonding. Interestingly, this is not seen for Eu deposition on either the $TiO_2$/MgO or YSZ[12] at elevated temperatures. At this point, while the in-plane rotation, in conjunction with the $TiO_2$ RHEED pattern and lack of re-evaporation, would suggest that we have successfully reduced the interfacial energy at the interface by limiting electrostatic effects, one possibility that cannot be ruled out is Eu-Ti-O reactivity at the interface and that the lack of re-evaporation is due to some complex composition.

To investigate the magnetic properties of 5nm EuO/$TiO_2$(1.5ML)/MgO(10nm)/MgO(001), the magneto-optic Kerr effect (MOKE) is measured in an optical flow cryostat with variable temperature control. Longitudinal MOKE was measured with a p-polarized 635nm diode laser and an incident angle near 45 degrees with respect to an applied in-plane magnetic field (H). Figure 3 (E) inset shows representative M-H hysteresis loops at T = 6 K with a coercivity ($H_c$) of 117 Oe and ratio ($M_r$/$M_s$) between magnetization remanence ($M_r$) and saturation ($M_s$) of 0.53. Representative loops at T = 60 K and T = 74 K are also shown. In Figure 3 (E), $M_r$ is plotted (in degrees) as a function of temperature. Starting at 6 K, the Kerr rotation at remanence is 0.19 degrees and decreases with increasing temperature, following typical Curie-Weiss behavior down to the transition temperature at 69 K, the bulk $T_C$ value for EuO.

We next investigate the interfacial structure and material quality at the interface between EuO and the 1.5 ML $TiO_2$/MgO stacking in the following manner. A $TiO_2$ layer is grown on an MgO buffer layer and maintained at 500°C. Next, a Eu flux is exposed to the heated $TiO_2$ *without* introducing an oxygen partial pressure. Unlike the case for Eu flux incident on the bare MgO, the RHEED pattern immediately changes (Fig. 4 (A) and 4 (B)), indicating bonding of Eu atoms to the TiO$_2$ surface. Furthermore, the faint streaks between the underlying MgO(001) [100] RHEED pattern (Fig 4(A)) indicates layer-by-layer epitaxial growth of EuO(001) [110] // MgO(001) [100] in the ultrathin limit. As in the case of oxygen-free growth of EuO on YSZ(001),[12] the oxygen atoms are believed to be supplied by the substrate. After a few minutes, the RHEED pattern stabilizes indicating steady state re-evaporation of the incoming Eu flux and thus the growth is terminated. The short time frame and visible underlying MgO RHEED pattern suggests that at most, only a few monolayers of material are deposited.

MOKE measurements on the ultrathin EuO layer are shown in Fig 4 (C). Hysteresis loops taken at 6 K (Fig 4 (C) inset) clearly show ferromagnetic behavior with $H_c$ = 98 Oe and $M_r/M_s$ = 0.24. A temperature dependence of the magnetization remanence shows the transition temperature to be 69 K, indicating that the initial growth mode for Eu flux incident on the TiO$_2$/MgO interface is EuO and not a reacted Eu-Ti-O compound. Interestingly, the fact that $T_C$ is equal to the bulk value suggests that the resulting film thickness is large enough to avoid finite size effects, which should decrease $T_C$.[28] Furthermore, these magnetic results shed light on the initial growth mode seen in the RHEED time evolution (Fig 3 (C)). The reconstruction streaks immediately fade once the Eu flux is incident upon the TiO$_2$ surface. This occurs because the TiO$_2$ layer minimizes the electrostatic interactions between the EuO and MgO layers and creates nucleation sites for subsequent EuO epitaxy. The ability for Eu atoms to find a favorable binding site in the 2x2 reconstructed TiO (see Fig 1 (D)), results in the formation of EuO with oxygen supplied by the substrate.

4. Conclusion

In conclusion, electrostatic interactions at the interface between EuO and MgO can greatly determine the growth sequence of the EuO layer. To improve the epitaxy of EuO on MgO, a $TiO_2$ interfacial template was introduced and shown to alleviate like-ion repulsion and decrease the structural mismatch between EuO and MgO. Furthermore, the initial growth sequence is drastically different with the $TiO_2$ interface than on the bare MgO as demonstrated by in-plane rotation and fast strain relaxation. Also, the addition of the $TiO_2$ layer allows for substrate-supplied oxidation leading to ultrathin ferromagnetic EuO films. Such a template could be an avenue for combining emerging materials onto MgO such as $EuTiO_3$ or other rock salt magnetic oxides in single crystal heterostructures

Acknowledgements

We would like to acknowledge stimulating discussions and technical assistance from Kathy McCreary, Wei Han, and Pat Odenthal, and funding from DARPA/Defense Microelectronics Activity (DMEA) under agreement number H94003-10-2-1004.

Figure Captions

Fig. 1. (Color online) (A) Schematic of cube-on-cube EuO/MgO(001) in a 4:5 (EuO:MgO) configuration at the interface. Ions are represented as follows: the Mg ions are shown as small spheres (orange), the O in MgO ions are large white spheres, Eu ions are the medium spheres (blue) and O in EuO are the large dark spheres (green). (B) Shows the ion size effect for cube-on-cube growth of EuO (transparent over layer) on MgO (under layer). (C) Configuration for a 45° rotated EuO over layer on MgO demonstrating the anion-anion overlap between the oxygen ions of the EuO and MgO. (D) Structure of the EuO/TiO$_2$/MgO layers. Ti

ions are the smallest gray spheres and O ions in TiO$_2$ are the large gray spheres (red). Boxes show the unit cells for each oxide. For (A) - (D), the ions in each schematic are sized according to their ionic radius. The MgO and TiO$_2$ are drawn to scale with the bulk MgO lattice parameter while all EuO layers correspond to the bulk EuO lattice constant.

Fig. 2. RHEED patterns for the 10 nm MgO buffer layer in the (A) MgO(001) [100] and (B) MgO(001) [110] directions. (C) and (D) are the RHEED patterns for 1 ML TiO$_2$ on MgO(001) in the [100] and the [110] directions, respectively. (E) and (F) are the RHEED patterns for 1.5 ML TiO$_2$ in the [100] and the [110] directions. (G) and (H) are the RHEED patterns for 2 ML TiO$_2$ in the [100] and the [110] directions. (I) and (J) are RHEED patterns for a 5 nm EuO films on the 2 ML TiO$_2$ layer for EuO(001) [110] // MgO(001) [100] and EuO(001) [100] // MgO(001) [110], respectively.

Fig. 3. (A) and (B) are the RHEED patterns for a 5 nm EuO thin film deposited on TiO$_2$(1.5ML)/MgO(10nm)/MgO(001) for EuO(001) [110] // MgO(001) [100] and EuO(001) [100] // MgO(001) [110], respectively. (C) is the time evolution of the EuO growth on TiO$_2$(1.5ML)/MgO(10nm)/MgO(001) over the initial 10 minutes of EuO growth The initial and final line cuts are shown above and below, respectively. In (C), the peaks in the initial line cut correspond to diffraction rods seen in the 1.5 ML TiO$_2$ RHEED pattern as seen in Fig. 2 (F), while the final line cut corresponds to the diffraction rods seen for EuO [100] // MgO [110] as seen

in Fig. 3 (B). (C1) (dashed line) indicates when the Eu flux is incident on the $TiO_2$ layer and (C2) (dashed line) indicates the introduction of $O_2$ into the system. (D) The time evolution of direct deposition of EuO on MgO(10nm)/MgO(001) in the MgO [110] direction and the peaks in the initial line cut shown above correspond to the diffraction rods as shown in Fig. 2 (B). (D1) (dashed line) indicates when the Eu flux is incident on the MgO and (D2) (dashed line) indicates the introduction of $O_2$. Below (D) is the final line cut of EuO after 30 minutes of growth directly on the MgO buffer layer. (E) Temperature dependence of the measured MOKE angle (plotted in degrees) taken at 0 Oe (remanence) for EuO (5nm)/$TiO_2$/MgO(001). Insert shows representative hysteresis loops for T = 6 K (Black), T = 60 K (Red or grey) and T = 74 K (Blue or dark grey)

Fig. 4. RHEED patterns for Eu deposition on the $TiO_2$ layer without leaking $O_2$ into the system in the (A) MgO(001) [100] and (B) MgO(001) [110] directions. (C) Temperature dependence of the MOKE signal measured at saturation and the insert shows a representative hysteresis loop at T = 6 K.

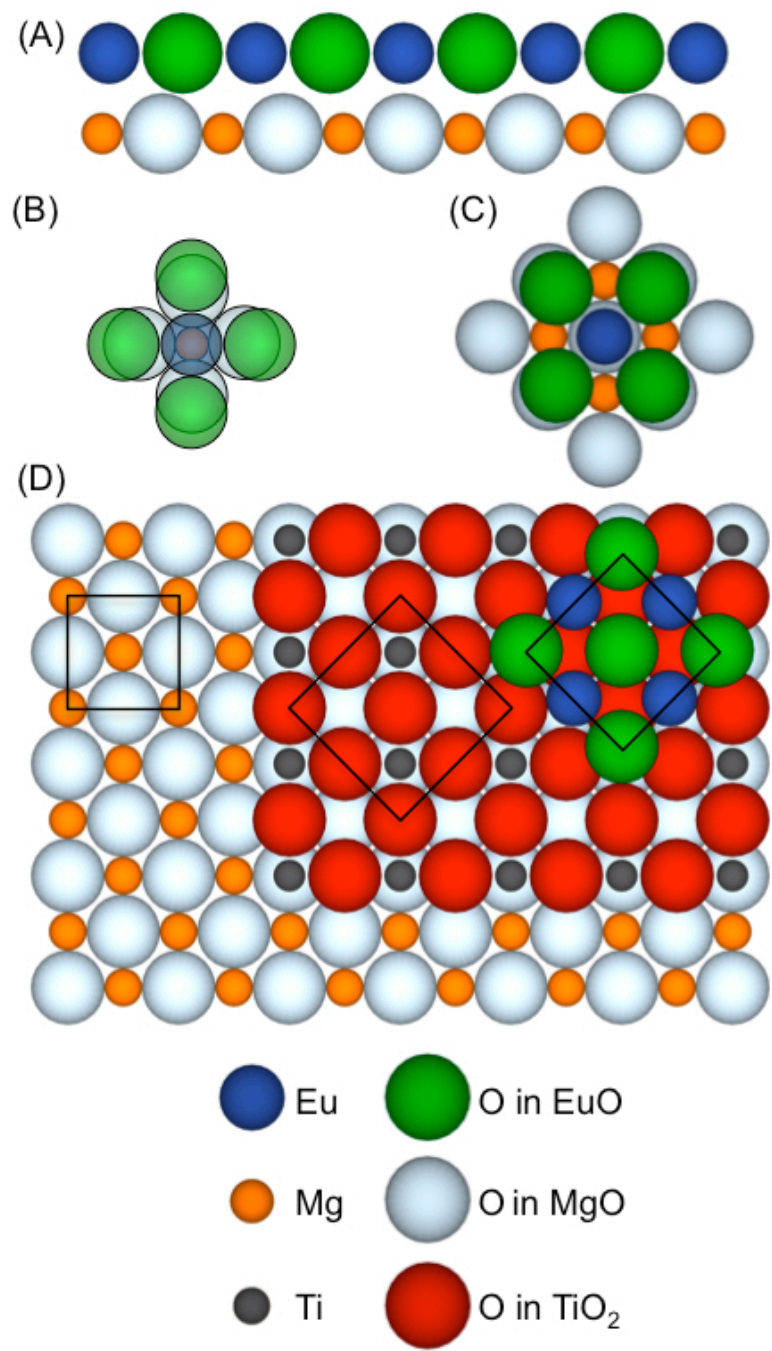

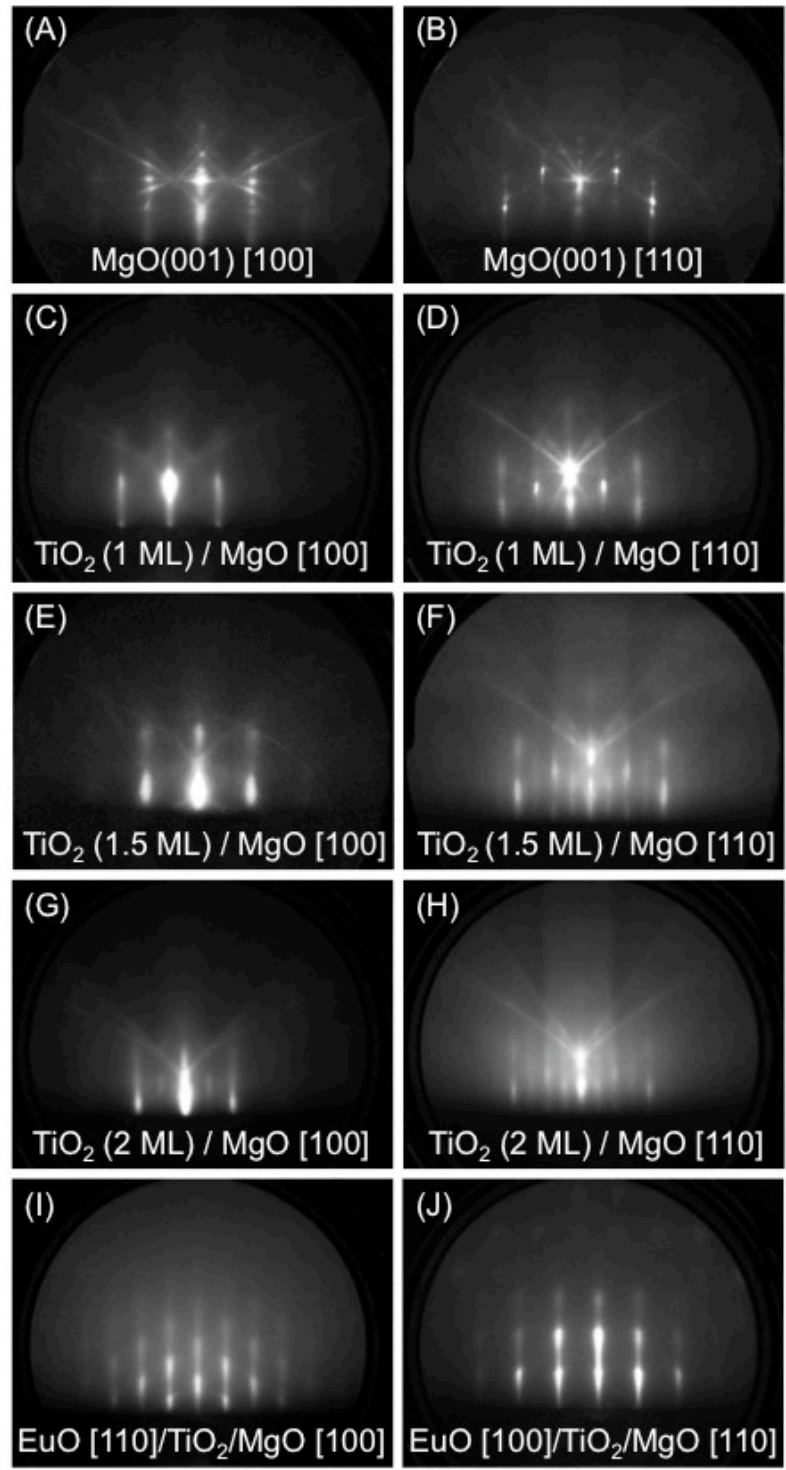

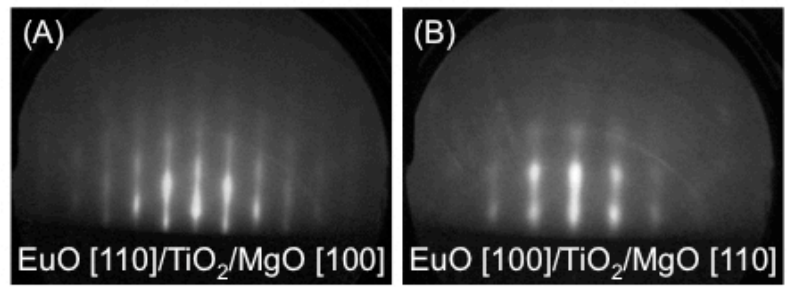

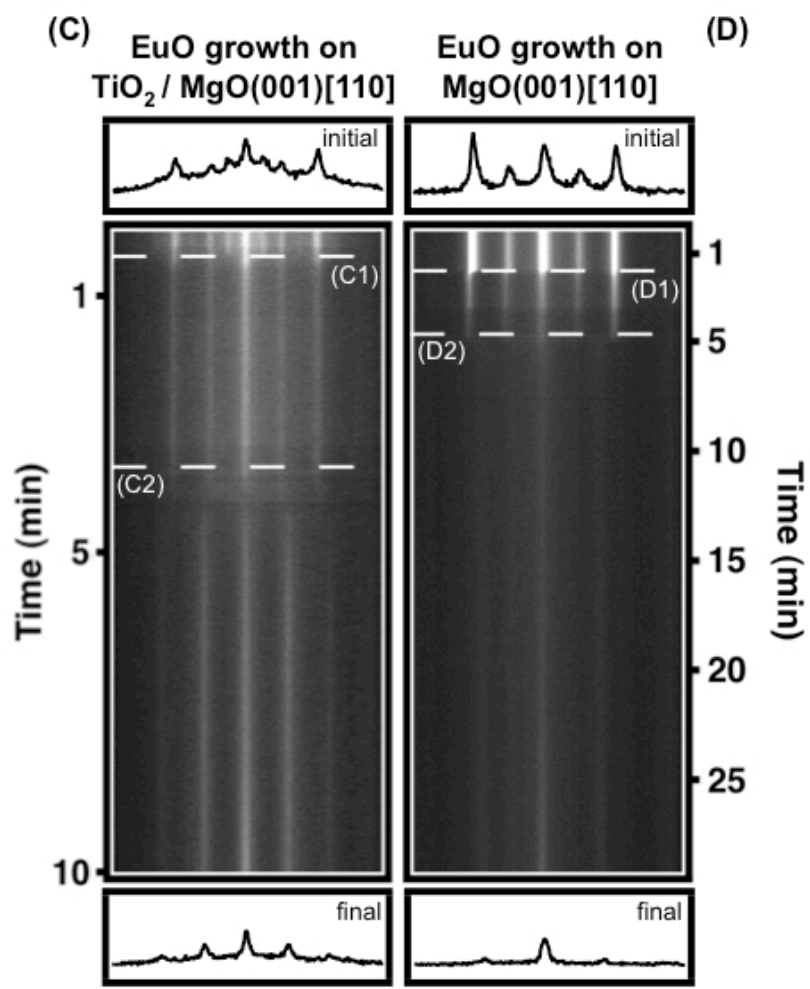

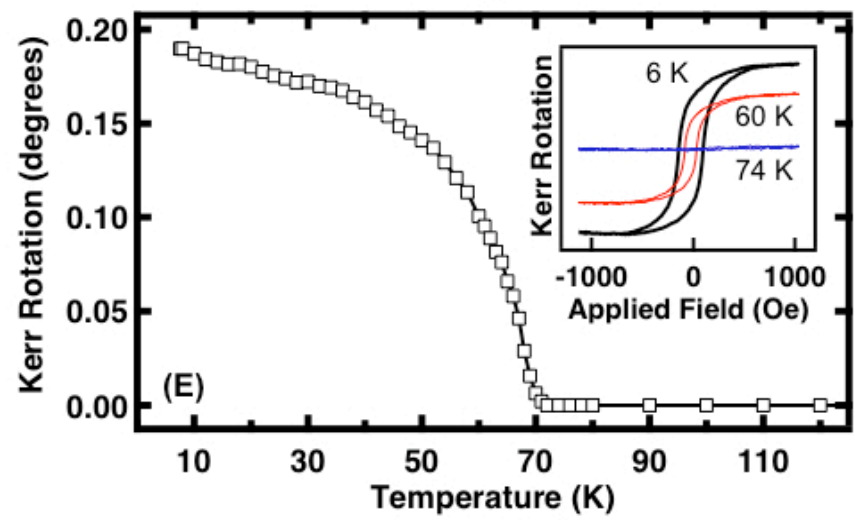

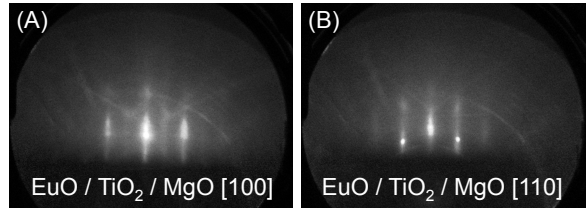
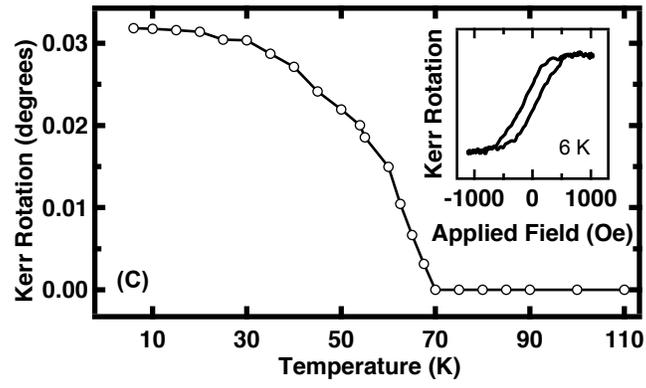